\documentclass[]{elsarticle}

\usepackage{lineno,hyperref}
\modulolinenumbers[5]

\journal{Journal of Computers and Fluids}

\usepackage{caption}
\usepackage{subcaption}
\usepackage{amsmath}
\usepackage{xcolor}
\usepackage{svg}
\begin{document}

\begin{frontmatter}

\title{Data-Driven Multi-Grid Solver for Accelerated \\ Pressure Projection}

%% Group authors per affiliation:
\author{Gabriel D Weymouth}
\address{Engineering and Physical Sciences, University of Southampton, Southampton, UK}
\address{Data-Centric Engineering, Alan Turing Institute, London, UK}
\ead[url]{https://weymouth.github.io/}

\begin{abstract}
Pressure projection is the single most computationally expensive step in an unsteady incompressible fluid simulation. \textcolor{black}{This work demonstrates the ability of data-driven methods to accelerate the approximate solution of the Poisson equation at the heart of pressure projection.} Geometric Multi-Grid methods are identified as linear convolutional encoder-decoder networks and a data-driven smoother is developed using automatic differentiation to optimize the velocity-divergence projection. The new method is found to accelerate classic Multi-Grid methods by a factor of two to three with no loss of accuracy on eleven 2D and 3D flow cases including cases with dynamic immersed solid boundaries. The optimal parameters are found to transfer nearly 100\% effectiveness as the resolution is increased, providing a robust approach for accelerated pressure projection of unsteady flows.
\end{abstract}


\end{frontmatter}

\section{Introduction}

Pressure projection is a bottleneck in high-speed unsteady incompressible flow solvers. Constructing approximate solutions for the discrete pressure Poisson system is the most expensive part of each time-step as the elliptic equation requires communication throughout the computational domain instead of being confined to a local characteristic. As such, the proportional cost only grows in massively parallel simulations due to the required communication across processes. Methods such as artificial-compressibility \cite{he2002comparison}, smooth-particle-hydrodynamics \cite{kiara2013sph}, and particle-in-cell \cite{jiang2017angular} attempt to side-step this computational cost by modelling or otherwise under-resolving the pressure evolution compared to the fluid's momentum. However, these approaches lead to significant errors in pressure forces, thereby making them unsuitable for many applications or requiring explicit pressure corrections, \cite{kiara2013sph}.

\begin{figure}
    \centering
    \includegraphics[width=0.8\textwidth]{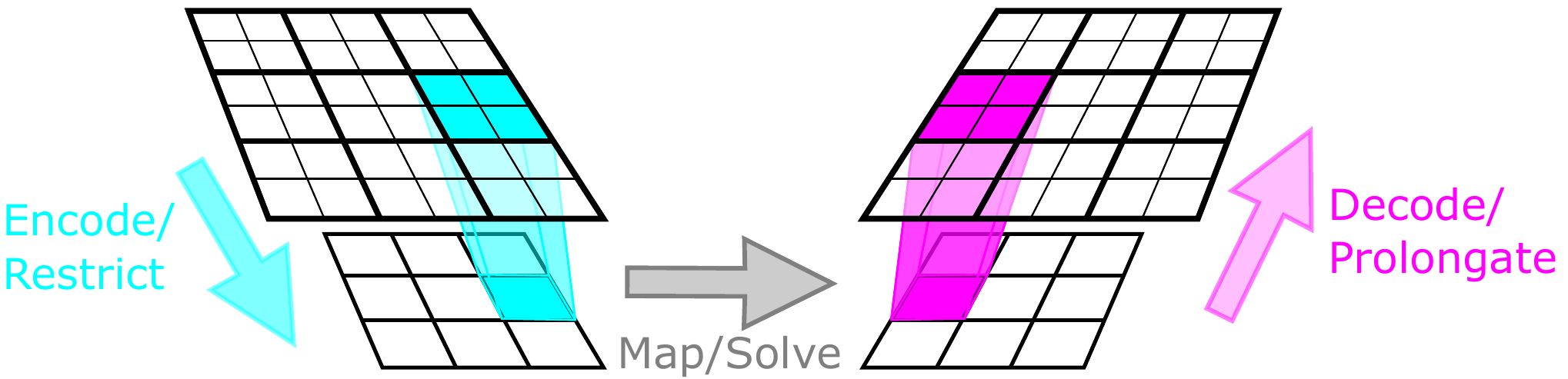}
    \caption{Sketch of a convolutional encoder-decoder network and/or a geometric multi-grid V-cycle. Data is restricted down to a reduced size for faster processing before being pushed back up to the full size. A two-level V-cycle is shown, but this process repeats recursively between the largest and smallest grid levels.}
    \label{fig:multigrid}
\end{figure}

Advances in data-driven acceleration of flow solvers have mainly focused on reducing simulation resolution using data-assimilation and data-driven turbulence closures \cite{asch2016data,BECK2019108910,maulik_san_rasheed_vedula_2019,ling_kurzawski_templeton_2016,font2021deep} without focusing on projection explicitly. Many of these approaches utilize Convolutional Neural Networks (CNN), Figure \ref{fig:multigrid}, which have been successfully applied in other fluids applications such as super-resolution \cite{liu2020deep} and full field regression models \cite{bhatnagar2019prediction}. The CNN's inherent translation symmetry and the encoder-decoder architecture's reduction and expansion of the array size both help constrain model learning capacity and generalize the trained networks to unseen prediction cases. 

A few recent studies have applied machine-learning to accelerate the critical projection step itself \cite{ozbay2021poisson,Xiao2020,ajuria2020} using CNNs to predict the pressure field given the projection source term. All of these method assume a uniform grid without internal solid geometries, although \cite{ozbay2021poisson} uses a decomposition to handle different \textit{domain} boundary conditions and grid aspect ratios in 2D. Unfortunately, the methods can't be applied to most simulation problems and even state-of-the-art results are still qualitative, with errors greater than 10\% throughout much of the domain \cite{ozbay2021poisson}. 

A key insight to accelerating pressure projection is that classic Geometric Multi-Grid (GMG) methods \cite{briggs2000multigrid} are simply \textit{linear} convolutional encoder-decoder networks, Figure \ref{fig:multigrid}. While there is a long history of optimizing multi-grid methods (both geometric \cite{tang2000sparse} and algebraic \cite{brezina2006adaptive}), few of these have taken a direct data-driven approach. \textcolor{black}{In \cite{tang2000sparse}, a sparse approximate smoother is constructed by solving a series of least-square matrix equations instead of a data-driven approach. This improved the convergence properties on extremely skewed grids compared to Gauss-Sidel smoothers, but not on more regular grids or when using a very sparse (and therefore fast) approximate inverse.} 

Employing modern data-driven techniques could  further accelerate GMG without the loss of accuracy currently limiting CNN methods.  Recent work has shown promising initial results in constructing data-driven restriction and prolongation operators \cite{KATRUTSA2020112524,greenfeld2019learning}. \textcolor{black}{However, improving the architecture of these operators is a very challenging discrete optimization problem, limiting the results achieved thus far. Both \cite{KATRUTSA2020112524,greenfeld2019learning} show improved convergence spectra compared to classic GMG methods, but no overall speed-up. }

\textcolor{black}{Building on this foundation, this manuscript develops a novel data-driven smoother which is optimized using automatic differentiation to minimize the velocity-divergence over the full recursive GMG V-cycle. This new framework achieves 2 to 3-fold acceleration compared to standard GMG on eleven 2D and 3D benchmark projection problems including those derived from unsteady Navier-Stokes simulations with dynamic immersed boundaries.} Moreover, since the new approach maintains linearity in the pressure and the nonlinear dependence on the matrix coefficients are scale-invariant, the data-optimized parameters generalize extremely well to new flows and simulation resolutions.

\section{Linear System Description}

The discrete Poisson equation in the projection step is defined as
\begin{equation}\label{eq:axb}
    A x = b
\end{equation}
where $x$ is the unknown pressure field vector, $b$ is the source term proportional to the divergence of the velocity field to be projected, and $A$ is the discrete matrix for the Poisson operator. For a conservative Poisson operator, $A$ is symmetric, has zero-sum rows and columns, is negative semi-definite, and has a single zero eigenvalue. \textcolor{black}{The zero eigenvalue means the solution is unique up to an additive constant, in agreement with the derivative action of the pressure on the velocity-divergence. While it is possible to add an additional condition to the matrix, this is not required and has no impact on the projection problem or the measured pressure forces.} The matrix is also extremely sparse. While the solution vector size $N$ may easily be $10^6-10^8$, a second-order scheme on structured grids in $M$ dimensions result in only $M$ non-zero sub-diagonals.

Iterative methods solve equation \ref{eq:axb} by updating an approximate solution $x^k$ to a new solution $x^{k+1}$ with reduced error. As the problem is linear, the equation for the update $\epsilon$ is simply
\begin{equation}\label{eq:aer} 
    A \epsilon^k = r^k \equiv b - Ax^k
\end{equation}
where $r$ is the residual. \textcolor{black}{In the context of the pressure Poisson equation, the residual is the remaining divergence in the velocity field. Driving this residual to zero by updating the pressure is the goal of the projection step.} 

In practise, only an approximate solution for $\epsilon$ is obtained, after which the solution and residual are incremented
\begin{equation}\label{eq:increment}
    x^{k+1} = x^k+\epsilon^k, \quad r^{k+1} = r^k-A\epsilon^k.
\end{equation}
\textcolor{black}{This process is iterated until the velocity-divergence residual is reduced to a specified tolerance, at which point the projection step is complete.}

Geometric Multi-Grid (GMG) methods are among the fastest iterative solution approaches for variable coefficient discrete pressure Poisson equations. A single iteration of GMG is called a V-cycle and consists of (i) a solution precondition step, (ii) residual restriction down to the reduced-size problem, (iii) approximate solution of this problem, (iv) solution prolongation back up to correct the fine grid, and (v) solution smoothing. This process is recursive, being applied to the reduced-size problems until they are small enough to be solved directly, resulting in $O(N)$ operations per V-cycle.
The first four steps of the V-cycle distribute the residual throughout the domain, which enables simple and relatively fast stationary methods such as Gauss-Sidel or Successive Over Relation (SOR) to effectively smooth the local error in the projected solution. The V-cycle can be repeated until convergence or used as an efficient preconditioner for another iterative solver such as Conjugate Gradient. 

\section{Data-driven Accelerated Projection Method}

As the Poisson equation is elliptic, \textit{any} element of the velocity-divergence residual $r$ potentially influences \textit{every} element of the update $\epsilon$. Therefore the linear $O(N)$ scaling achieved by GMG is already optimal. However, data-driven methods can still be used to accelerate GMG projection by speeding-up and increasing the residual reduction of each V-cycle iteration.

The linearity of equation \ref{eq:aer} is a critical property which the data-driven projection method must maintain. Without it, the projection method can not be applied iteratively to drive the velocity-divergence to the required tolerance of the flow solver. However, this linearity is only with respect to the update and residual fields. The GMG operators can potentially be explicit nonlinear functions of $A$, embedding information about the discrete problem and boundary conditions. 

Which GMG operation is the best candidate for such an approach? Significant effort has focused on the optimization of the prolongation operator and its transpose restriction operator. However, prolongation/restriction operators are not explicit functions of $A$, making their optimization mathematically and numerically complex \cite{KATRUTSA2020112524,greenfeld2019learning}. And while CNN encoding/decoding are simple to optimize with back-propagation, their nonlinear activation functions make them invalid in this application. 

Another option is the precondition operator, but a simple Jacobi preconditioner
\begin{equation}
    \epsilon = D^{-1}r
\end{equation}
where $D$ is the diagonal of $A$, is sufficient for the V-cycle to distribute the residual throughout the domain and is as fast as possible; being a single Schur (broadcasted) vector product. Therefore, for the remainder of the paper we will use a simple Jacobi preconditioner and uniform pooling/distribution for restriction/prolongation and focus on a data-driven smoothing operator.

While more complex options are certainly possible, this work considers a sparse approximate inverse as a smoother
\begin{equation}\label{eq:smooth}
    \epsilon = \tilde A^{-1}r
\end{equation}
where $\tilde A^{-1}=f(A\,|\theta)$ is a parameterized pseudo-inverse with the same sparsity as $A$ and $\theta$ is the parameter vector. The advantage of this pseudo-inverse smoother is speed. Unlike Gauss-Sidel or SOR smoothers, the sparse matrix-vector multiplication of equation \ref{eq:smooth} requires no back-propagation and so can be vectorized in serial or parallel implementations, offering a significant potential speed-up. Additionally, the matrix $A$ is constant during the projection step (and often for an entire simulation), meaning $\tilde A^{-1}$ can be computed and stored ahead of time.

There are few constraints on $\tilde A^{-1}$: it must scale with $A^{-1}$, and symmetry of $A$ implies the pseudo-inverse should also be symmetric. For simplicity, the diagonal and off-diagonal coefficients of $\tilde A^{-1}$ are constructed independently 
\begin{equation}\label{eq:approxinv}
    \tilde a^{-1}_{ii} = \frac{f_d(a_{ii}/s\,|\theta_d)}{a_{ii}} , \quad
    \tilde a^{-1}_{ij} = \frac{f_o(a_{ij}/s\,|\theta_o)}{a_{ii}+a_{jj}}
\end{equation}
where function inputs are scaled by the maximum off-diagonal $s=\max(A-D)$ and the outputs are scaled by the diagonal elements. Note that $f_o(0)=0$ is required to maintain the sparsity of $\tilde A^{-1}$ and that a Jacobi smoother would use a diagonal function $f_d=1$ and off-diagonal function $f_o=0$. With the optimization problem now reduced to two normalized single-variable functions, the specific choice of parameterization is not critical and a simple quadratic is chosen for $f_d$ and $f_o$. Using higher-order polynomials, splines, and interpolating kernels did not significantly change the results.

The parameterized smoother is optimized for a set of training data $X=\{A,r^0\}$. The loss is defined as the reduction in the log-$L_2$-norm of the residual after each V-cycle iteration 
\begin{equation}
L = \log_{10}(|r^{k+1}|_2 / |r^{k}|_2)
\end{equation}
This loss is averaged over each iteration and each example in the data-set. \textcolor{black}{Working directly with a residual loss has a number of advantages. First and foremost, the purpose of the pressure projection step is to drive the velocity-divergence residual to zero, making this the ultimate judge of any projection method. Second, this incorporates problem-specific residual data which would be missing from a metric based only on the V-cycle operator. Finally, the residual is always available, unlike the error with respect to the unknown exact pressure solution.}

The optimal parameters of the accelerated GMG projection method are then given by the minimization
\begin{equation}
    \hat\theta = \min_\theta L(\theta\, |X)
\end{equation}
The residual loss function is a highly nonlinear recursive function of the parameters, but automatic differentiation (AD) is used to determine the gradient $\nabla_\theta L$ and Hessian $H_{\theta\theta} L$ in a Newton's method optimizer  \cite{mogensen2018optim,RevelsLubinPapamarkou2016}. The use of AD allows this data-driven solver to extend to an arbitrary number of grid levels, avoiding inaccurate and potentially unstable finite differences.

The data-driven solver and all test cases presented in the following sections are implemented in the Julia programming language \cite{bezanson2017julia} and available for reproduction \cite{weymouth2021julia}.

\begin{figure}
    \centering
    \includegraphics[width=0.8\textwidth]{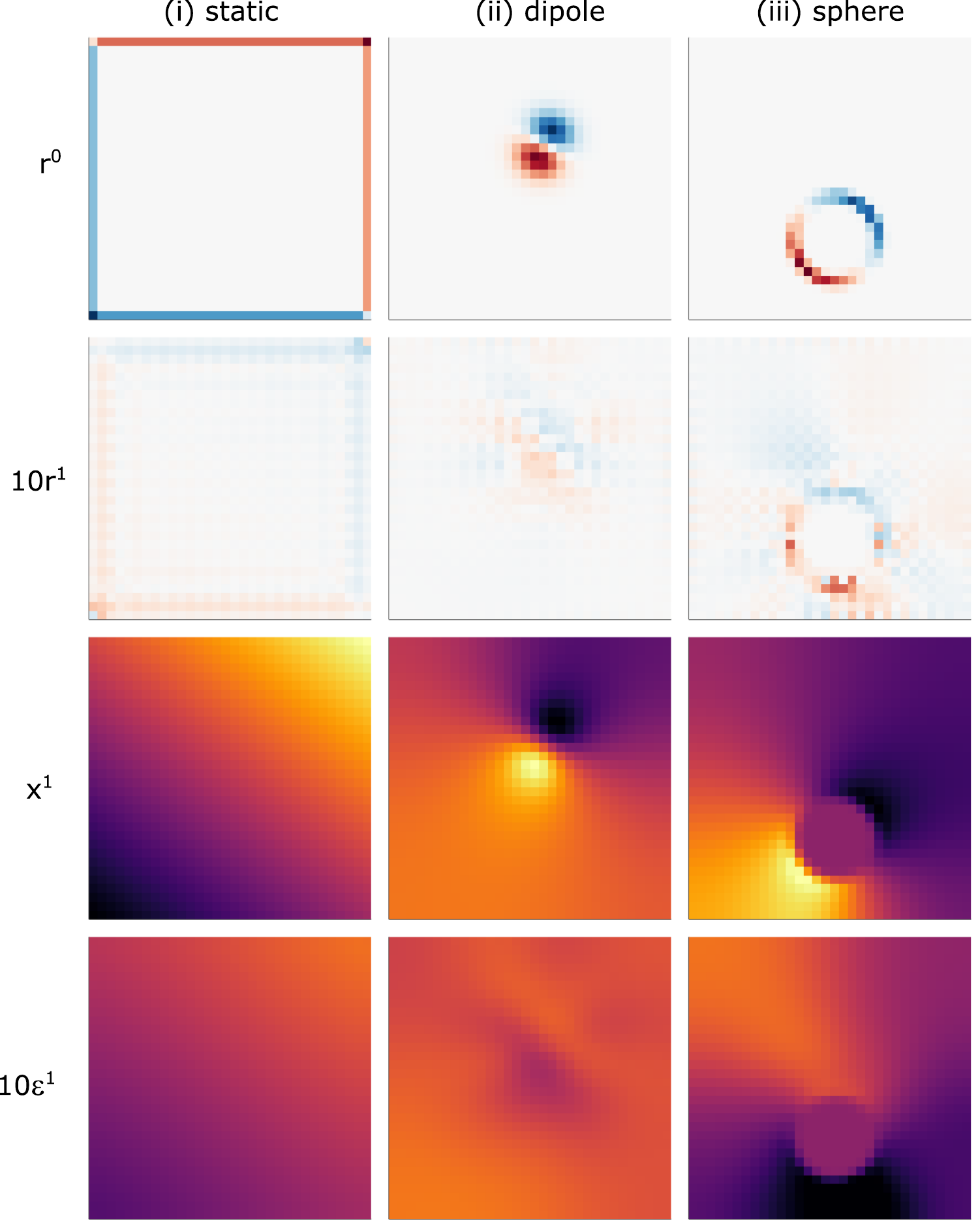}
    \caption{\textcolor{black}{Examples from the 2D static (left), dipole (center), and sphere (right) synthetic cases. The initial residual $r^0$ (row-1) and 10x residual after one data-driven V-cycle $10r^1$ (row-2) plotted on the same scale. The solution after one data-driven V-cycle $x^1$ (row-3) and it's 10x error compared to a machine precision solution $10\epsilon^1$ (row-4) plotted on the same scale.}}
    \label{fig:synthetic cases}
\end{figure}

\section{Synthetic Discrete Poisson System Results}

\textcolor{black}{A set of six synthetic cases were developed to establish the ability of the data-driven approach to project out residuals (i) on domain boundaries, (ii) within the fluid, or (iii) on body boundaries. The three 2D cases and their highly localized residuals are shown in Figure~\ref{fig:synthetic cases}, and each has a matching 3D case. The `static' cases feature a random linear pressure field generated by residuals on the domain boundaries (i). The `dipole' and `sphere' cases feature residuals on a random dipole in the fluid (ii) or on the boundary of a random immersed sphere (iii). Neumann conditions are applied on the domain and sphere boundaries using the Boundary Data Immersion Method (BDIM) to adjust the $A$ matrix coefficients, as described and validated in \cite{maertens2015accurate,Lauber2022}. The  generative equations for each synthetic case are given in \ref{app synth}.}

A data-driven GMG solver is optimized for each synthetic case using 100 randomized training examples. \textcolor{black}{Figure~\ref{fig:synthetic cases} also shows the residual and error of each case after one data-driven V-cycle. The results show that the initially highly-localized residual is projected throughout the domain, without build-up of error on the domain or immersed boundaries. Repeated V-cycles drives the residual and error to machine zero in every case.} 

\begin{figure}
    \centering
    \begin{subfigure}[b]{0.47\textwidth}
        \centering
        \includegraphics[width=\textwidth]{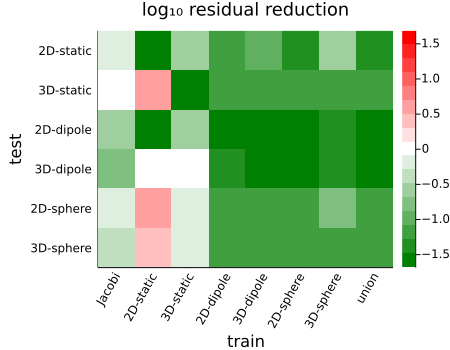}
        \caption{single-cycle residual reduction}
        \label{fig:cross plot}
    \end{subfigure}
    \hfill
    \begin{subfigure}[b]{0.5\textwidth}
        \centering
        \includegraphics[width=\textwidth]{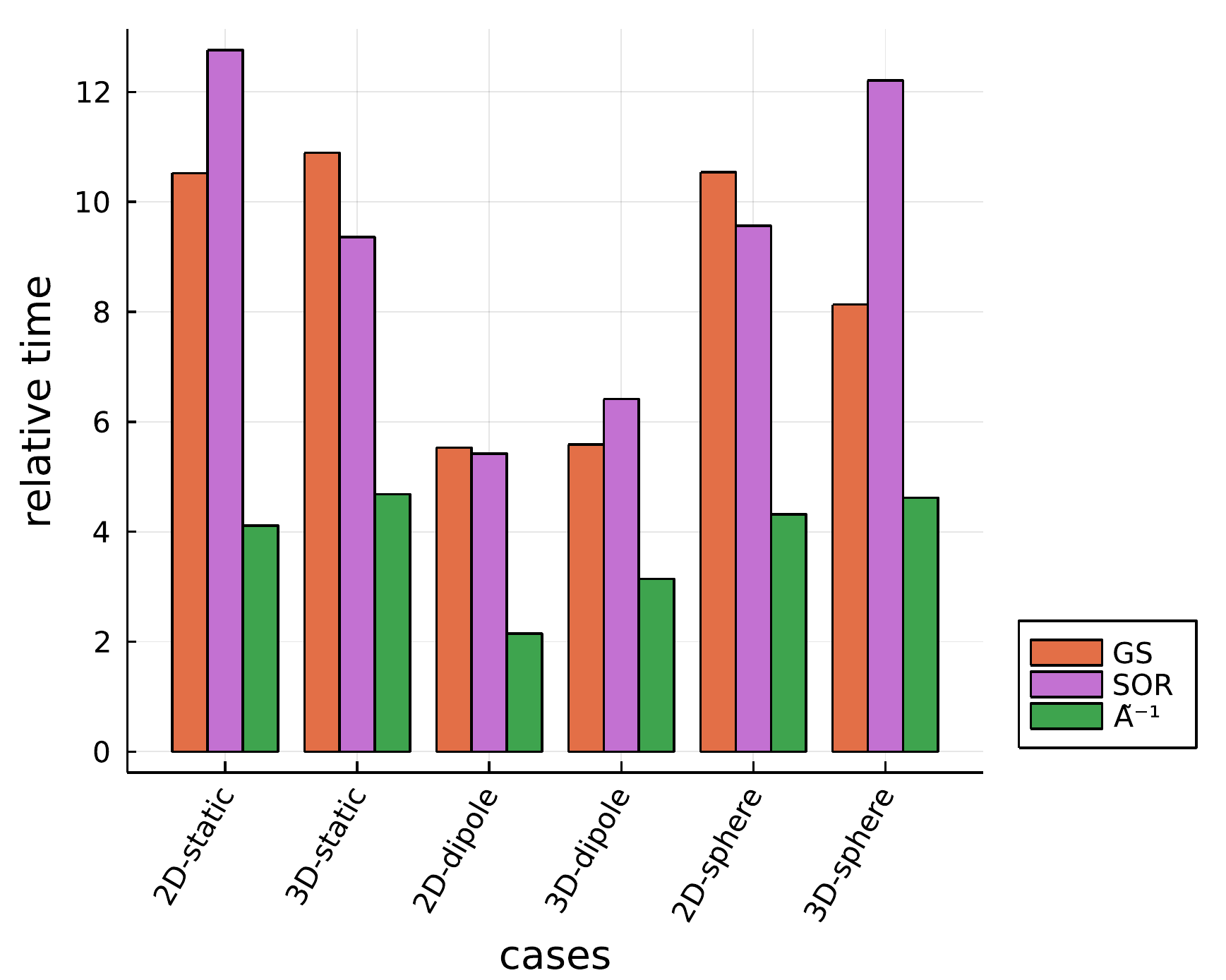}
        \caption{solver time}
        \label{fig:synthetic time}
    \end{subfigure}
    \caption{a) Residual reduction factor $|r_{k+1}|/|r_k|$ over a single Multi-grid V-cycle on the `test' case after tuning using the `train' case. `Jacobi` is the V-cycle performance using a classic Jacobi smoother, and `union' refers to the solver trained on all of the synthetic cases. (b) Time to reduce pressure residual by $10^{-3}$ for classical and parameterized smoothers on each synthetic case. Time is relative to a single Jacobi-smoothed V-cycle on each grid.}
    \label{fig:synthetic results}
\end{figure}

Figure~\ref{fig:cross plot} quantifies the generalization performance of each trained projection method on 100 unseen test examples across cases. \textcolor{black}{The results for a Jacobi-smoothed GMG V-cycle are shown for comparison as this is the initial condition for the \textit{untrained} data-driven method.} After training, the data-driven method vastly outperform the Jacobi V-cycle, doubling the residual loss on the 3D-dipole case and providing nearly 30-fold improvement on the 3D-static case. While the performance of the data-driven projection method is best when testing and training on the same case, most of the solvers still improve on the Jacobi-smoothed GMG performance for cases outside of their training set. Indeed, the solver trained on the 2D-sphere generalizes essentially as well as a solver trained on the `union' of the all training case data. The exceptions are the static cases, with the solver trained on the 2D-static case failing to converge on three of the other test cases. Note that despite identical $A$ matrices in the static and dipole cases, the performance of the those solvers are completely different. This indicates that the data-driven methods are specializing based on the residual data as well as the matrix structure, a feature unique to this approach.

\textcolor{black}{Finally, the acceleration of the data-driven projection method is evaluated against high-performance GMG solvers using classic stationary smoothers; Gauss-Sidel (GS) and Successive-Over-Relaxation (SOR).} The `union'-trained data-driven solver is used and 100 tests are generated for each case on different grids to test the ability of the data-driven method to generalize to new grid sizes.

The results in Figure~\ref{fig:synthetic time} show that the data-driven method has no issue projecting the velocity divergence on unseen grid sizes, reducing the residual by $10^{-3}$ in only 2-5x the time of a single Jacobi-smoothed V-cycle. \textcolor{black}{This is chosen as our unit of time since Jacobi-smoothing requires the smallest possible number of operations per V-cycle, normalizing results across grids.} As the pseudo-inverse smoother is many times faster than the Gauss-Sidel and SOR smoothers, this results in an overall 80-210\% speed-up (mean: 133\%) relative to classic GMG solvers. Note that this speed-up will be even more favorable in parallel computation as the new smoother operates directly on the local residual without back-propagation, eliminating any additional communication.

\begin{figure}
    \centering
    \begin{subfigure}[a]{\textwidth}
        \centering 
        \includegraphics[width=\textwidth]{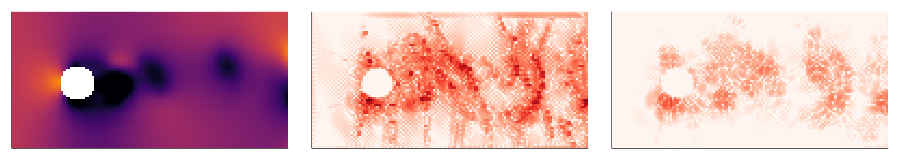}
        \caption[-1mm]{Circle flow}
    \end{subfigure}
    \begin{subfigure}[b]{\textwidth}
    
        \centering
        \includegraphics[width=\textwidth]{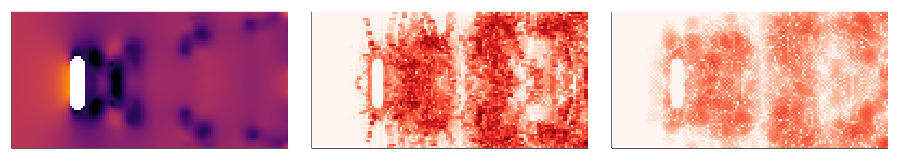}
        \caption{Slice through 3D donut flow}
    \end{subfigure}
    \begin{subfigure}[c]{\textwidth}
        \centering
        \includegraphics[width=\textwidth,trim={0 150 0 0},clip=true]{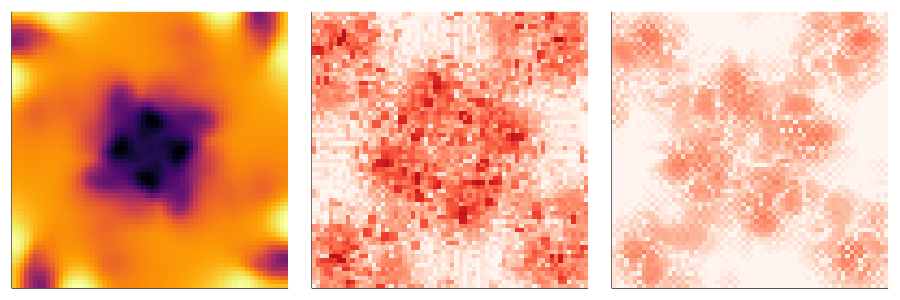}
        \caption{Slice through 3D Taylor-Green Vortex (TGV)}
    \end{subfigure}
    \begin{subfigure}[c]{\textwidth}
        \centering
        \includegraphics[width=\textwidth,trim={0 100 0 60},clip=true]{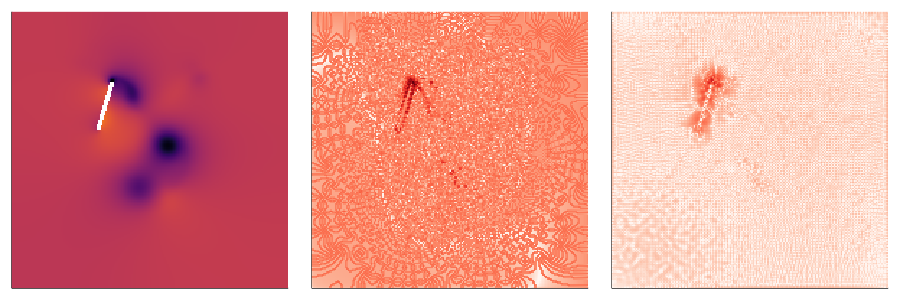}
        \caption{Flapping wing}
    \end{subfigure}
    \begin{subfigure}[b]{\textwidth}
        \centering
        \includegraphics[width=\textwidth]{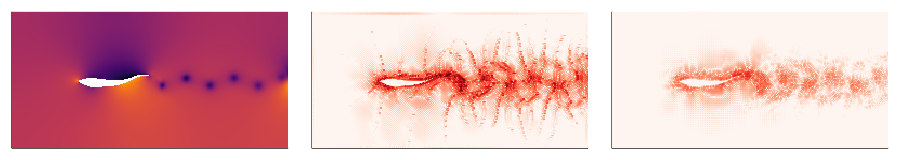}
        \caption{Swimming shark}
    \end{subfigure}
    \hfill
    \caption{Snapshots from the five flow simulation cases showing the pressure solution (left), the initial residual $|r^0|$ (middle), and after a single V-cycle $|r^1|$ (right) using the data-driven projection method. The TGV and donut flow are 3D, the flapping wing is dynamic without a background flow and the swimming shark is a deforming geometry. All residuals are visualized on the same log-scale, $10^{-6} \ldots 10^{-1}$. The TGV and wing images have been zoomed in for display purposes.}
    \label{fig:simulation cases}
\end{figure}

\section{Unsteady Incompressible Simulation Projection Results}

Five unsteady incompressible simulation cases were used to further characterize the accelerated projection method, Figure \ref{fig:simulation cases}. \textcolor{black}{The first three cases are variations on standard unsteady flow benchmarks, while the second two are variations of recent validated biologically-inspired flows \cite{Lauber2022}.} The simulations feature significantly different physics to test the accelerated projection performance from flows with and without background velocities, immersed geometries, body motion, and body deformation, as well as using computational grids of different sizes and aspect ratios. \textcolor{black}{All simulations use the same extensively validated unsteady incompressible Navier-Stokes BDIM solver \cite{maertens2015accurate,Lauber2022} and 100 Poisson matrices $A$ and initial residuals $r^0$ are copied from each unsteady simulation to form the training and testing data. The details of each case are found in \ref{app flow}.}

\begin{figure}
    \centering
    \includegraphics[width=0.7\textwidth]{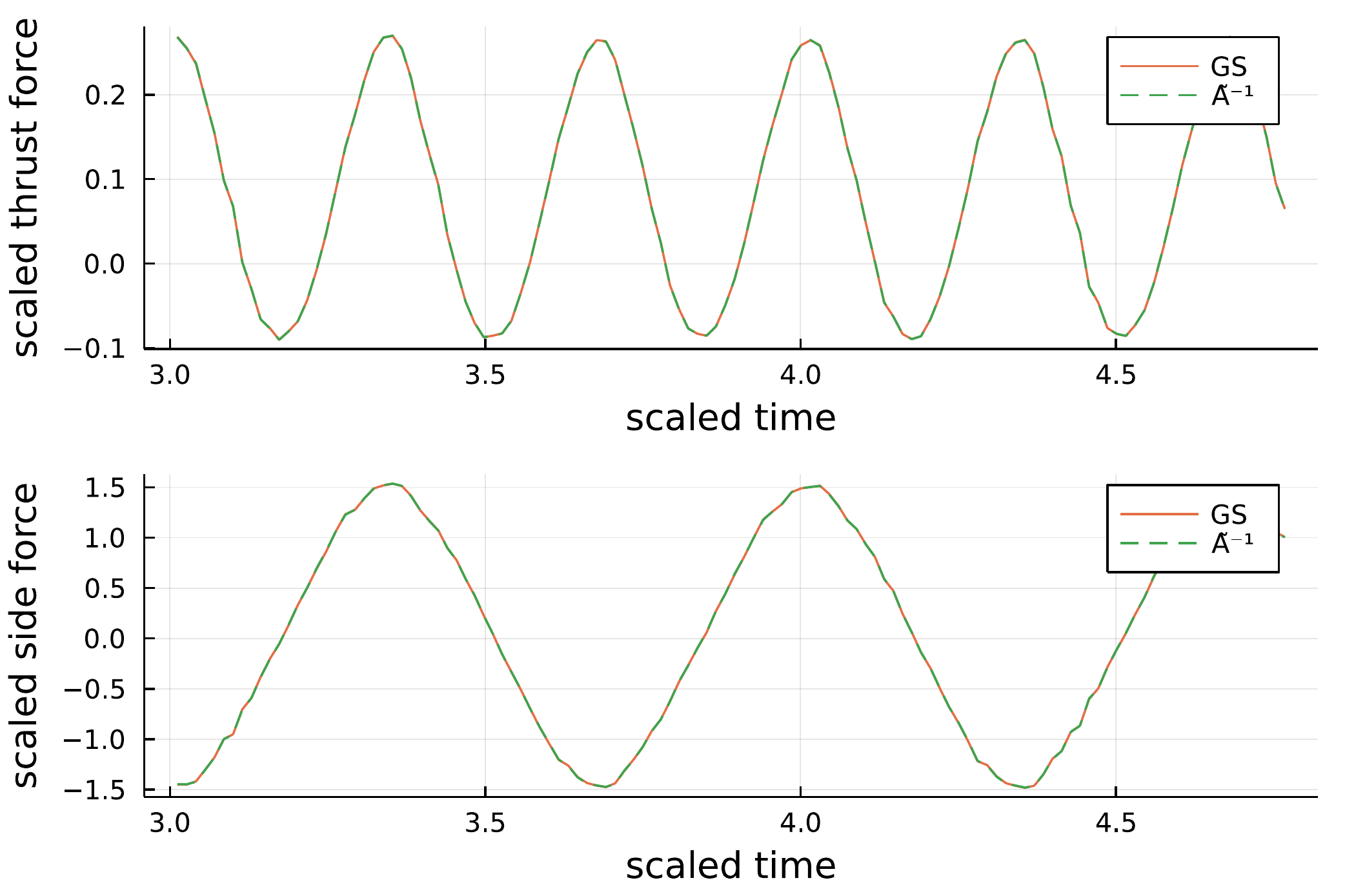}
    \caption{\textcolor{black}{Integrated thrust and side pressure forces the swimming shark test case using the Gauss-Sidel (GS) and data-driven ($\tilde A^{-1}$) GMG solvers and $|r^k/r^0|<10^{-3}$ tolerance level.}}
    \label{fig:simulation forces}
\end{figure}

Figure \ref{fig:simulation cases} shows the pressure solutions, initial residual field, and the residual after a single V-cycle of the data-driven GMG method trained on each case. The results show that the residual decreases throughout the domain, including at the domain and immersed boundaries. \textcolor{black}{This is further verified by Fig~\ref{fig:simulation forces}, which demonstrates that the unsteady swimming shark pressure forces using the data-driven GMG solver is indistinguishable from the Gauss-Sidel smoothed results.} As such, the new data-driven method is shown to produce uniformly valid solutions despite only being trained to optimize the global residual loss.

\begin{figure}
    \centering
    \begin{subfigure}[b]{0.52\textwidth}
        \centering
        \includegraphics[width=\textwidth]{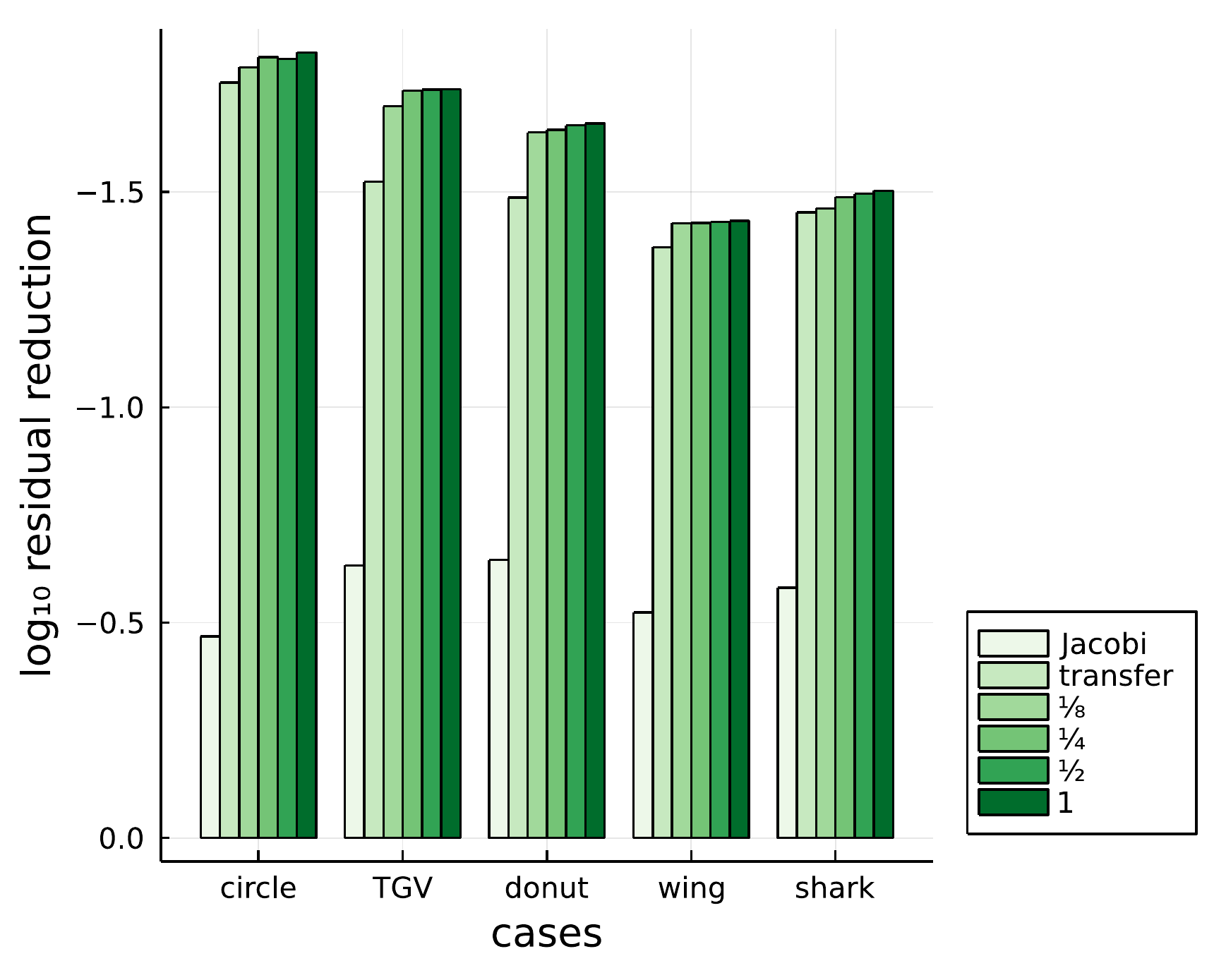}
        \caption{single-cycle residual reduction}
        \label{fig:scaled loss}
    \end{subfigure}
    \hfill
    \begin{subfigure}[b]{0.42\textwidth}
        \centering
        \includegraphics[width=\textwidth]{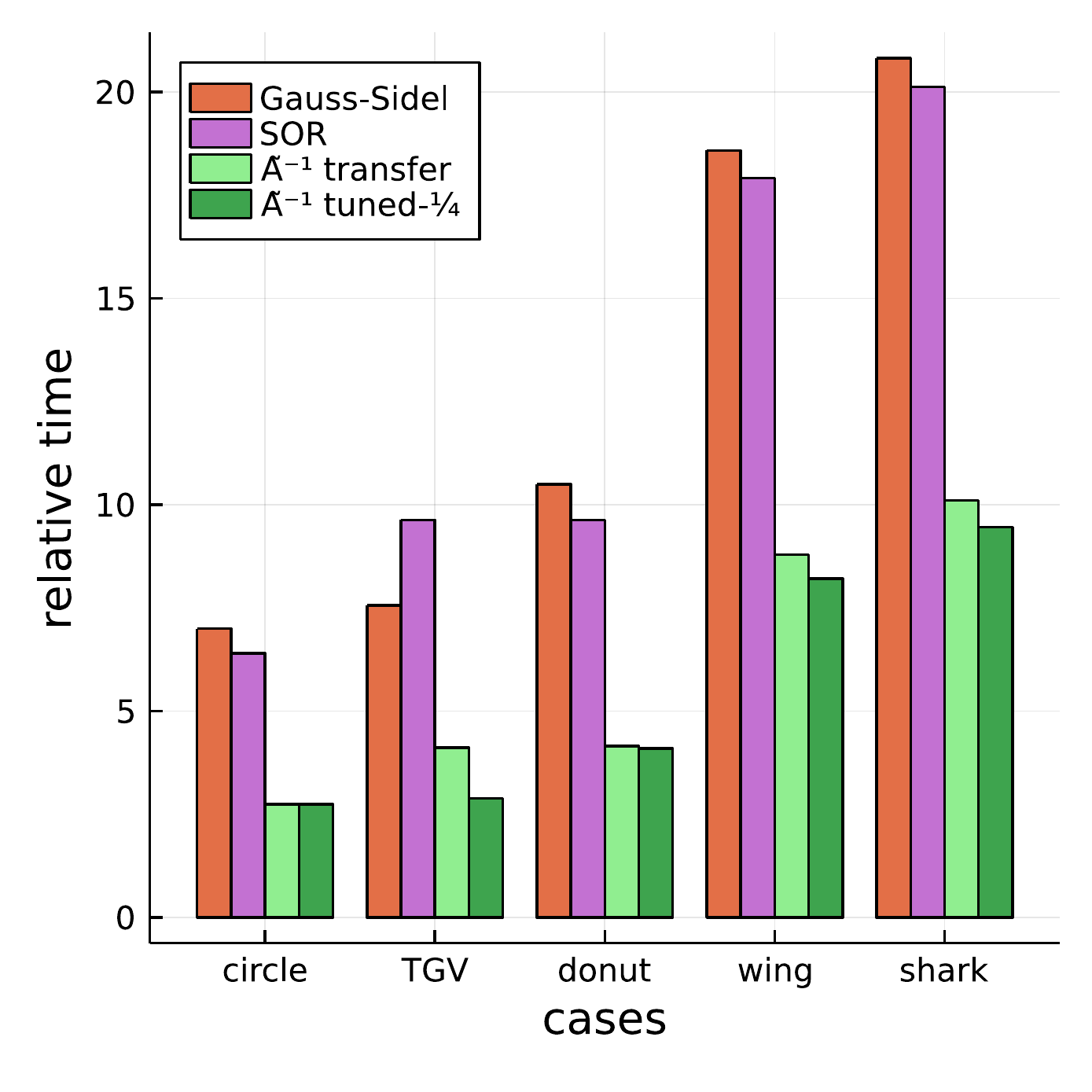}
        \caption{solver time}
        \label{fig:simulation time}
    \end{subfigure}
        \caption{(a) Residual reduction over a single Multi-grid V-cycle on the full-resolution case. The `transfer' smoother has been trained on the union of the synthetic data sets from Figure~\ref{fig:synthetic cases}. The $\frac 18, \frac 14, \frac 12$ smoothers have been trained on simulations with the indicated reduced resolution \textit{in each spacial and temporal dimension}. (b) Time to reduce pressure residual by $10^{-3}$ for classical and parameterized smoothers on each simulation case. Time is relative to the time of a single V-cycle using the Jacobi smoother.}
        \label{fig:tuned simulation}
\end{figure}

Figure \ref{fig:scaled loss} shows that when the data-driven method trained on the synthetic cases of the previous section is transferred to these unseen simulation cases, it reduces the residual 2.5 to 3.9 times more effectively than an untrained Jacobi-smoothed V-cycle. Indeed, without any further training this transfer solver achieves 85-95\% of the residual reduction of a smoother optimized for each new case. Even more promising is that training on reduced-size simulations closes that small gap almost completely. Training on a $1/4^\text{th}$-scale simulation is fast, requiring only $N/4^M$ points (N/16 in 2D and N/64 in 3D) and $\approx 1/4$ the number of time steps, while achieving 99-100\% of the residual reduction of training with full-scale data. Such an \textit{auto-tuning} approach enables a highly effective data-driven projection method to be developed for any flow simulation case with very little overhead. Figure \ref{fig:simulation time} shows that this data-driven method greatly accelerates the projection step relative to the classic MG methods on these real flow cases. The solver tuned on the reduced resolution cases provides 112 - 230\% acceleration, and even the transferred solver provides a mean acceleration of 120\%.

\begin{figure}
    \centering
    \begin{subfigure}[b]{0.47\textwidth}
        \centering
        \includegraphics[width=\textwidth]{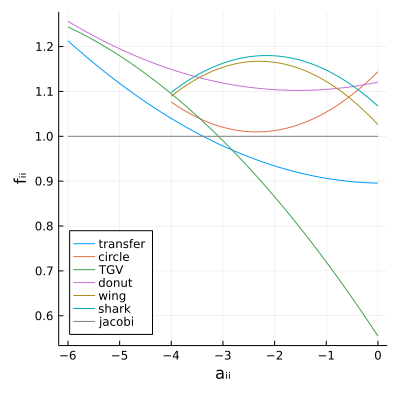}
        \caption{diagonal functions $f_d$}
    \end{subfigure}
    \hfill
    \begin{subfigure}[b]{0.47\textwidth}
        \centering
        \includegraphics[width=\textwidth]{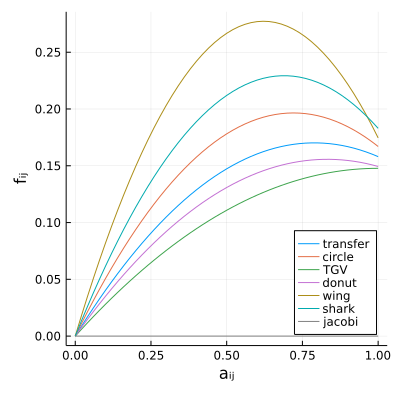}
        \caption{off-diagonal functions $f_o$}
    \end{subfigure}
        \caption{(a) Diagonal and (b) off-diagonal parameterized functions after optimization on the 1/4-resolution case. The `transfer' smoother was tuned on the synthetic cases and the Jacobi smoother values are shown for comparison.}
        \label{fig:tuned inverse}
\end{figure}

Finally, the parameterized $f_d,\,f_o$ functions used in the approximate inverse equation \ref{eq:approxinv} are shown in Figure \ref{fig:tuned inverse}. It is interesting to note that the diagonal functions are all somewhat centered on one, the value for the Jacobi smoother. We also note that the `wing' and `shark' cases produce parameterizations which are similar to each other but significantly different than the others cases. This is reasonable as dynamic and deforming geometries put unique burdens on the pressure projection step, as shown in the longer convergence times for these cases shown in Figure \ref{fig:simulation time}. Adopting a data-driven and auto-tuned approach enables these pressure-projection dominated cases to achieve significant accelerations.

\section{Conclusions}

This manuscript develops a successful data-driven method to accelerate the solution of discrete pressure Poisson systems found in incompressible flow simulations. Geometric Multi-Grid (GMG) methods are identified as linear convolutional encoder-decoder networks with optimal $O(N)$ scaling, and the matrix coefficients are identified as a critical nonlinear input, not only the projection source-term, as they embed information such as boundary conditions. Mathematical constraints are used to further focus the learning capacity to a parameterized sparse pseudo-inverse Multi-grid smoother. The resulting data-driven solver is within 33\% of the minimum computational cost per V-cycle, and shown to accelerate classic Gauss-Sidel and SOR smoothed GMG solvers by 80-233\% on eleven simulation cases. Because of the focused learning capacity, the generalization is excellent, enabling 90\% effective transfer learning from a synthetic data-set and nearly 100\% transfer from reduced resolution simulations.

The potential of machine learning advances to improve fluid dynamics is vast, but well-applied classical methods and constraints are needed to focus this potential. Wherever possible, this work has made the simplest choice in parameterization, leaving significant opportunities for future improvements. \textcolor{black}{The flow cases in this manuscript use Cartesian-grids, but this does not limit the generality of the projection problems as the resulting $A$ matrices are nonuniform due to the presence of the immersed geometries. The current data-driven GMG framework can therefore be readily extended to the nonuniform matrices induced by stretched structured grids. Extensions to unstructured grids would require the use of algebraic instead of geometric multi-grid, and a similar data-driven sparse smoother could accelerate such projection methods.}

\bibliography{mybibfile}
\color{black}
\appendix
\section{Synthetic case details}\label{app synth}

The synthetic cases are all generated with simple functions on uniform 2D or 3D grid using $n$ grid cells in each direction. $n=32$ cells were used for the training data and the testing data in Fig~\ref{fig:cross plot}. $n=64,96$ and $128$ were used for the testing data in Fig~\ref{fig:synthetic time}.

The 2D and 3D static cases are generated from a constant gradient solution
$$ x_i = \hat m \cdot \vec q_i$$
where $\vec q_i$ is the vector to the center of grid-cell $i$ and $\hat m$ is a random unit vector. The matrix $A$ is defined by $a_{ij}=1$ on cell faces, $a_{ij}=0$ on domain faces, and $a_{ii}=-\sum_j a_{ij}$. The initial residual is then $r^0=A x$ which is nonzero only on the domain boundaries.

The 2D and 3D dipole cases are generated using a localized residual function
$$ r_i = \hat m \cdot \left(\vec q_i-\vec Q\right)\ e^{-\frac{|\vec q_i-\vec Q|^2}{\sigma^2 }} $$
where $\vec Q$ is a random location in the fluid domain and $\sigma$ is a random width. The same uniform $A$ matrix is used as in the static case.

The 2D and 3D sphere cases are generated using the residual from an immersed sphere with random center $\vec Q$ and radius $s$, moving in a random direction $\hat m$ in an initially still fluid. Following the Boundary Data Immersion Method \cite{maertens2015accurate,Lauber2022} the conservative  pressure equation is then
$$ \oint_{\partial\Omega_i} \frac{\partial \phi}{\partial n}\beta \,\text{d}s = \oint_{\partial\Omega_i} m_n \,(1-\beta)  \,\text{d}s $$
where $\partial\Omega_i$ are the faces of grid cell $i$, $\hat n$ is the face normal, $\phi$ is the scaled pressure, and $\beta$ is a coefficient which smoothly transitions from 1 to 0 as a function of the signed distance $d$ from the grid cell face to the immersed boundary. For example, a sphere has signed distance $d=|\vec q-\vec Q|-s$, and we could use $\beta=\min(1,\max(0,d+\frac 1 2))$.

This equation is applied on each unit cell using second order central-finite difference and cell-center quadrature. For example, in 1D this gives
\begin{gather*}
a_{i,i-1}=\beta^w_i,\ a_{i,i+1}=\beta^e_i ,\ a_{ii}=-(\beta^w_i+\beta^e_i),\ a_{ij}=0\ \text{else}\\
r_i =  -m\left(\beta^e_i-\beta^w_i\right)
\end{gather*}
where $\beta^{w,e}_i$ are the values on the west and east faces of cell $i$. The residual is non-zero only in the immersed boundary transition region where $\beta^e\ne\beta^w$.

\section{Unsteady simulation case details}\label{app flow}

The details for the five unsteady flow cases shown in Figure~\ref{fig:simulation cases} are:
\begin{enumerate}[(a)]
    \item 2D flow past a static circular cylinder: Radius $r=32$ grid cells, domain size $16r$ by $8r$ cells, cylinder center $(4r,4r)$, and Reynolds number $\text{Re}=Ur/\nu=250$.
    \item 3D flow past a static donut: Major radius $R=32$ cells, minor radius $r=R/4$, domain size $8R$ by $4R$ by $4R$, center $(2R,2R,2R)$, and $\text{Re} = RU/\nu=10^3$.
    \item 3D Taylor-Green Vortex turbulent decay: initial velocity function $u=-\sin(kx)\cos(ky)\cos(kz)$, $v=\cos(kx)\cos(ky)\cos(kz)$, $w=0$ with $k=\pi/n$, domain size $n=128$ cubed, and $\text{Re}=10^5$.
    \item 2D flow induced by a flapping plate wing: Length $L=64$ cells, initial position $(3L,4L)$, horizontal displacement $x=L\sin(\omega t)$, rotation angle $\alpha=\frac\pi 4\cos(\omega t)$, domain size $6L$ by $6L$ cells, and $\text{Re}=\omega L^2/\nu=250$.
    \item 2D flow past the undulating cross-section of a swimming shark: Body length $L=128$ cells, domain size $4L$ by $2L$, undulation wavenumber $k=5.3/L$, tail peak-to-peak amplitude $A=0.2L$, Strouhal number $\text{St}=fA/U=0.3$, and $\text{Re}=UL/\nu=10^4$.
\end{enumerate}
Reduced resolution training data used for Figure~\ref{fig:tuned simulation} uses the same setup as above, except the resolution is decreased. For example, a $\frac 14$-resolution wing simulation uses $L=16$ cells which reduces the amplitude of motion and domain size proportionally.

\end{document}